\documentclass[journal=jpcbfk,manuscript=article,layout=twocolumn]{achemso}

\usepackage{amsmath}

\author{Hitoshi Ohta}
\email{hohta@kobe-u.ac.jp}
\affiliation
{Molecular Photoscience Research Center, Kobe University, Kobe 657-8501, Japan}
\alsoaffiliation
{Graduate School of Science, Kobe University, Kobe 657-8501, Japan}
\author{Takahiro Sakurai}
\affiliation
{Center for Supports to Research and Education Activities, Kobe University, Kobe 657-8501, Japan}
\author{Ryosuke Matsui}
\author{Kohei Kawasaki}
\author{Yuki Hirao}
\affiliation{Graduate School of Science, Kobe University, Kobe 657-8501, Japan}
\author{Susumu Okubo}
\affiliation
{Molecular Photoscience Research Center, Kobe University, Kobe 657-8501, Japan}
\author{Kazuyuki Matsubayashi}
\affiliation
{Institute for Solid State Physics, University of Tokyo, Kashiwa 277-8581, Japan}
\author{Yoshiya Uwatoko}
\affiliation
{Institute for Solid State Physics, University of Tokyo, Kashiwa 277-8581, Japan}
\author{Kazutaka Kudo}
\affiliation
{Department of Physics, Okayama University, Okayama 700-8530, Japan}
\author{Yoji Koike}
\affiliation
{Department of Applied Physics, Tohoku University, Sendai 980-8579, Japan}

\title{Frequency Extension to THz Range in High Pressure ESR System and Its Application to Shastry-Sutherland Model Compound SrCu$_{2}$(BO$_{3}$)$_{2}$}


\begin{document}








\begin{abstract}
We have made a survey of ceramics for the inner parts of the transmission-type pressure cell to achieve the high pressure and the high transmission in THz range.
By using the optimal combination of ZrO$_{2}$-based ceramic and Al$_{2}$O$_{3}$ ceramic, we have succeeded in obtaining the pressure up to 1.5 GPa and the frequency region up to 700 GHz simultaneously.
We show the high-pressure ESR results of Shastry-Sutherland compound SrCu$_{2}$(BO$_{3}$)$_{2}$ as an application.
We observed the direct ESR transition modes between the singlet ground state and the triplet excited states up to the pressure of 1.51 GPa successfully, and obtained the precise pressure dependence of the gap energy.
The gap energy is directly proved to be suppressed by the pressure.
Moreover, we found that the system approaches the quantum critical point with pressure by comparing the obtained data with the theory.
This result also shows the usefulness of high-pressure ESR measurement in THz region to study quantum spin systems.
\end{abstract}

\section{Introduction}
Pressure has been recognized as one of the most important parameters in study of condensed matter physics.
The pressure effect is studied by a variety of method such as electrical resistivity, magnetic susceptibility, specific heat, X-ray or neutron diffraction, NMR, inelastic neutron scattering and so on. Among them high-pressure ESR is useful to investigate the pressure effect of the magnetic material from the microscopic point of view.
Most of the reported high-pressure ESR systems \cite{ref15} except our and N\'afr\'adi's \cite{nafradi} systems are equipped with resonator and only one frequency is available in principal, while the sensitivity is higher than our systems.
Exceptionally, a method using the combination of the resonator and the plastic diamond anvil cell has been developed very recently \cite{hill}.
In its case, the multifrequency measurement can be done by using the higher resonant modes in addition to the fundamental mode.
However, the frequency region is limited in the 40-160 GHz frequency region, though the pressure reaches above 2 GPa.
N\'afr\'adi {\it et al.} whose system is a similar type with us started their development later than us and they have the system with the maximum field up to 16 T, the pressure region up to 1.6 GPa and the frequency region from 105 to 420 GHz \cite{nafradi}.

On the other hand, in our systems the frequency can be varied very widely, which includs the so-called THz region defined as the frequency region from 0.1 to 10 THz \cite{THz}, since we simply combine the piston-cylinder pressure cells with the transmission-type high-field ESR system using the pulsed high magnetic field \cite{ref1,ref2,ref3,ref4,ref5,ref6,ref7,ref8,ref9,ref10,ref11,ref12,ref13,ref14} or the static magnetic field \cite{ref15,ref16,ref17,ref18,ref19}.
The most characteristic point of our pressure cell is that all inner parts are made of sapphire and ZrO$_{2}$-based ceramic.
They have both the toughness and the transmission in the THz region, and this enables us to observe the transmitted light through the pressure cell in the wide frequency region.
The energy of the electromagnetic wave in this frequency region matches well the energy between the ground state and the low-lying excited states of the quantum spin systems.
Therefore, our high-pressure ESR systems have proved to be one of the most powerful means to study quantum spin systems \cite{ref1,ref2,ref3,ref4,ref5,ref6,ref7,ref8,ref9,ref10,ref11,ref12,ref13,ref14,ref15,ref16,ref17,ref18,ref19}.
So far, we developed the pressure cell with the moderate pressure range below 1 GPa and the wide frequency region from 70 to 700 GHz \cite{ref1,ref2,ref3,ref4,ref5,ref6,ref7,ref8,ref9,ref10,ref11}, and that with the higher pressure range up to 1.5 GPa and the limited frequency region from 70 to 400 GHz \cite{ref12,ref13,ref14,ref15,ref16,ref17}. 
Recently, we also developed a hybrid-type pressure cell with larger sample space and higher pressure region for large bore magnet, which covers the frequency region form 50 to 400 GHz and the pressure up to 2.5 GPa \cite{ref18,ref19}.
In order to achieve the high pressure and the wide frequency region simultaneously, the key factors are the toughness and the transmission property of the inner parts of the pressure cell.
In this paper, we describe the survey of several ceramics for the inner parts.
On the basis of this survey we tried new ceramic for the inner parts and we succeeded in extending the frequency region up to 700 GHz for the hybrid-type pressure cell.

\begin{figure}[t]
\begin{center}
\includegraphics[width=1\linewidth]{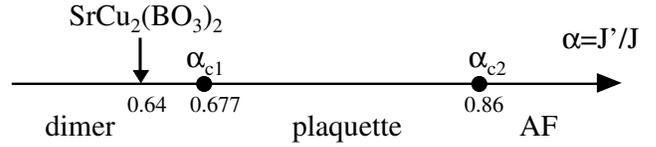}
\caption{\label{phase} Phase diagram for the Shastry-Sutherland model \cite{theory2}.}
\end{center}
\end{figure} 

As an application, we show the result of a two-dimensional spin-1/2 dimer system SrCu$_{2}$(BO$_{3}$)$_{2}$.
It is well known as a Shastry-Sutherland model compound which has an orthogonal dimer structure in two-dimensional plane \cite{kageyama}.
It shows a unique temperature dependence of the magnetic susceptibility and the ground state is the singlet.
From the fitting of this temperature dependence of the susceptibility with the theoretical calculation, the exchange interaction within dimer $J=$ 85 K and that between dimers $J'=$ 54 K were estimated \cite{JJ}.
Theoretically, the system is believed to be located near the phase boundary of the ratio $\alpha=J'/J$ between the interdimer and intradimer exchange interactions \cite{theory,theory2,theory3} as shown in Fig. 1.
The quantum critical point $\alpha_{c}$ was $\alpha_{c}\sim0.7$ and it was considered that there were only two phases of the dimer singlet phase ($\alpha<\alpha_{c}$) and the N\'eel phase ($\alpha>\alpha_{c}$) at the beginning \cite{theory}.
However, recent more detailed calculation for the Shastry-Sutherland model with the intradimer and interdimer exchange interactions by the series expansions has predicted that there exists an intermediate phase between these two phases \cite{theory2}.
It is the plaquette singlet phase ($\alpha_{c1}=0.677<\alpha<\alpha_{c2}=0.86$) as shown in Fig. \ref{phase}.
Moreover, the numerical exact diagonalization for the more realistic model with the interdimer Dzyaloshinsky-Moriya (DM) interaction in addition to these exchange interactions not only observed the anomaly in the $\alpha$ dependence of the gap energy around this predicted critical value $\alpha_{c1}$ but also explained the zero-field splitting of the direct ESR transition modes very well \cite{cepas}.
Since the system has the ratio $\alpha\sim0.64$ and it lies just below the critical point, it will provide an opportunity to study the quantum critical transition if the ratio can be controlled.
This has stimulated the study of the pressure effect on this system.
The magnetic susceptibility measurements up to 0.7 GPa suggested the suppression of the gap energy between the ground state and the first excited states with the pressure \cite{chi-T}, and we revealed the suppression of gap directly up to 1 GPa from the high-pressure ESR \cite{ref12}.
The NMR measurement at 2.4 GPa suggested the existence of the pressure-induced new phase where the magnetic and nonmagnetic sites coexist \cite{NMR}.
Moreover, the X-ray measurement suggested the complete collapse of the gap energy at 2 GPa and the quantum phase transition around at this pressure followed by the appearance of a new phase \cite{X-ray}.
However, the extrapolation of the pressure dependence of gap obtained from our previous ESR measurement up to 1 GPa suggested that the gap still remains at 2 GPa  \cite{ref12}.
This fact is pretty controversial with the X-ray study.
In this study, we extended the pressure range of the ESR measurement of SrCu$_{2}$(BO$_{3}$)$_{2}$ above 1.5 GPa successfully and we will show that the gap energy remains open certainly at 2 GPa.

\section{Experimental}
All measurements were performed by the recently developed ESR system using cryogen-free superconducting magnet \cite{ref18,ref19}.
The maximum magnetic field is 10 T.
Backward wave oscillators (BWO), which cover the frequency region from 300 to 800 GHz, are used as the light source.
The transmitted light trough the sample is detected by InSb detector set at the bottom of the light pipe.
The output of BWO is chopped mechanically with the frequency of several hundreds Hz and the output signal from the detector is amplified by the lock-in amplifier using the chopping frequency as the reference signal.
The detailed set up can be seen in ref. 21 and 22.

For the transmission measurement of the ceramics, we measured the transmission intensity of the sample by the above setup without field.
BWO is controlled by a PC through the AD converter and the frequency sweep measurement can be done.
The transmittance can be obtained by dividing the transmission intensity with sample by that without sample.
Although the output intensity of BWO has large frequency dependence, its reproducibility is high and the reliable transmittance can be obtained if the optical axis is maintained during the sample exchange. The sample ceramic has a cylindrical shape whose diameter and length are 12 mm and 30 mm, respectively. 
The length corresponds to the total length of the inner parts used in the pressure cell.
All transmission measurements have been done at 4.2 K and in the frequency region from 330 to 530 GHz.

The high-pressure ESR measurements of SrCu$_{2}$(BO$_{3}$)$_{2}$ have also been performed by the same setup.
The pressure cell including sample is connected to the light pipe so that the sample is located at the center of the magnet.
A rectangular shape single crystal $2\times2\times7$ mm$^{3}$ with its long axis along the $a$ axis was used.
The ESR measurements have been done at 2 K in the frequency region from 300 to 800 GHz.
The pressure cell is the hybrid-type piston-cylinder cell.
We used the inner parts made of the ZrO$_{2}$-based ceramic and the Al$_{2}$O$_{3}$ ceramic to achieve the transmission and the higher pressure simultaneously as is discribed later.
The maximum pressure was 1.51 GPa in this study.
The pressure was estimated from the load at room temperature by using the quadratic calibration curve of the load with the pressure around 3 K, which was obtained by the change of the superconducting transition temperature of tin.
The detail of the hybrid-type pressure cell can also be seen in ref. 21 and 22.

\section{Results and discussion}
\begin{table*}
\caption{Ceramics examined by the transmission measurement}
\label{tab1}       
\begin{tabular}{ccccc}
\hline\noalign{\smallskip}
product name&main component&$K_{1c}$&company&$b\times10^{3}$\\
&&(MPa$\cdot$m$^{1/2}$)&&(GHz$^{-1}$)\\
\noalign{\smallskip}\hline\noalign{\smallskip}
FCA10&Al$_{2}$O$_{3}$&3.1&Fuji Die Co. Ltd.&0.691\\
FCY40A&ZrO$_{2}$:Al$_{2}$O$_{3}$=60:40&5.3&Fuji Die Co. Ltd.&8.00\\
FCY20A\textsuperscript{\emph{a}}&ZrO$_{2}$:Al$_{2}$O$_{3}$=80:20&6.2&Fuji Die Co. Ltd.&11.1\\
FCY0M\textsuperscript{\emph{b}}&ZrO$_{2}$:Y$_{2}$O$_{3}$&7.1&Fuji Die Co. Ltd.&13.2\\
Z206N&ZrO$_{2}$:Y$_{2}$O$_{3}$=95:5&6$\sim$7&KYOCERA Corp.&7.90\\
Z201N&ZrO$_{2}$:Y$_{2}$O$_{3}$=93:7&4$\sim$5&KYOCERA Corp.&14.5\\
Z220&ZrO$_{2}$:MgO=95:5&7.8&KYOCERA Corp.&43.5\\
\noalign{\smallskip}\hline
\end{tabular}\\
\textsuperscript{\emph{a}} Currently used.;
\textsuperscript{\emph{b}} The ratio of the component is not disclosed.
\end{table*}

\subsection{Survey of ceramics for the inner parts of the pressure cell}
\begin{figure}[t]
\begin{center}
\includegraphics[width=0.9\linewidth]{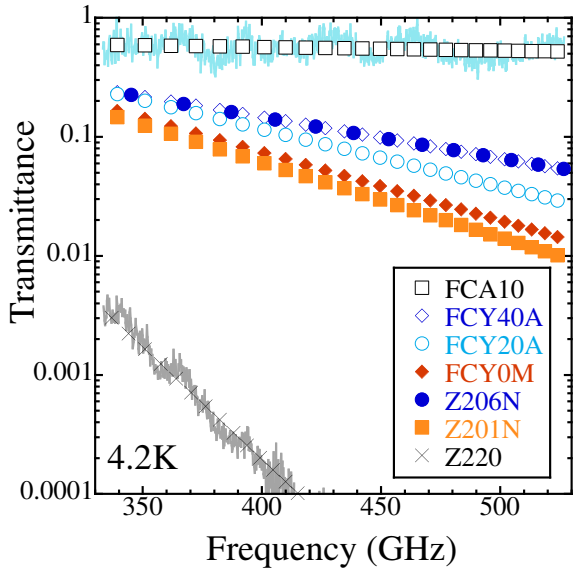}
\caption{\label{fig1} Frequency dependence of the transmittance of several ceramics.
For simplicity the raw data are shown only for FCA10 and Z220.
The open symbols indicate the ZrO$_{2}$ and Al$_{2}$O$_{3}$ based ceramics, the solid symbols and the cross symbol indicate the ZrO$_{2}$ ceramics including Y$_{2}$O$_{3}$ and MgO as a mixture, respectively.
The details of the examined ceramics are summarized in Table \ref{tab1}.}
\end{center}
\end{figure} 
\begin{figure}[t]
\begin{center}
\includegraphics[width=1\linewidth]{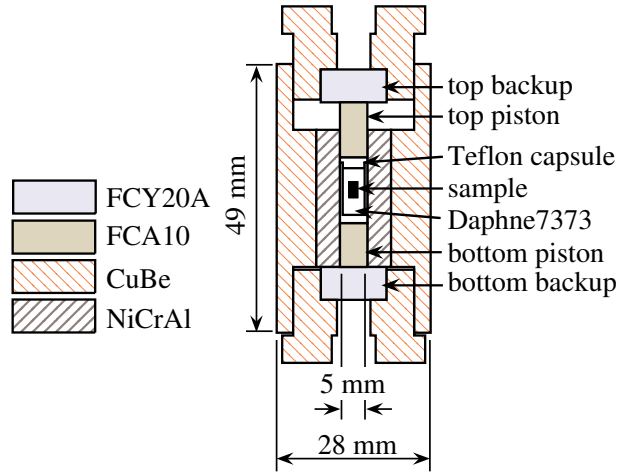}
\caption{\label{pcell} Cross section of the hybrid-type pressure cell \cite{ref18,ref19} with the inner parts setup for simultaneous achievement of the high pressure and the high transmission. Sample is set into a Teflon capsule and it is filled with the pressure-transmitting oil (Daphne 7373 in this study).}
\end{center}
\end{figure} 

The requirements for the inner parts of the pressure cell for our ESR system are the transmission in the THz region, the toughness, the absence of the magnetic impurity, the commercial availability, and so on.
Previously, sapphire and the ZrO$_{2}$-based ceramic FCY20A were used mainly as the inner parts material.
The sapphire has very high transmission and we can obtain the wide frequency region up to 700 GHz. 
However, it does not have enough toughness and it is easily damaged. When all inner parts are made of sapphire, the maximum pressure is 0.3 GPa at most.
And it does not reach 1 GPa even if several parts are replaced by that made of the tougher ZrO$_{2}$-based ceramic \cite{ref1,ref2,ref3,ref4,ref5,ref6,ref7,ref8,ref9,ref10,ref11}.
On the other hand, the ZrO$_{2}$-based ceramic has very high toughness and it is much cheaper than sapphire.
However, the ZrO$_{2}$-based ceramic does not have enough transmission especially for the higher frequency region \cite{ref12,ref13,ref14,ref15,ref16,ref17,ref18,ref19}.
In order to achieve the higher pressure and the wider frequency measurement simultaneously, finding a new material for the inner parts is an important factor.

The ZrO$_{2}$-based ceramic, which is known as the partially stabilized zirconia, is believed to have particular fracture toughness as compared with other ceramics in general.
Table \ref{tab1} shows the examined ceramics in this study.
All ceramics except FCA10 are the ZrO$_{2}$-based ceramic and FCA10 is made of only Al$_{2}$O$_{3}$. 
In this Table the main components, their contents and the fracture toughness values $K_{1c}$ are also shown, which are extracted from their data sheets.
Figure \ref{fig1} shows the frequency dependence of the transmittance of these ceramics.
All data are fitted well by a function $ae^{-b\nu}$, where $a$ and $b$ are the fitting parameter, and $\nu$ is the frequency.
The obtained parameter $b$ is also listed in Table \ref{tab1}.
As is seen clearly in Fig. \ref{fig1} and the obtained parameter $b$ in Table \ref{tab1}, the transmittance has large component dependence.
First, we focus on the difference of the transmittance for the series of FCA10, FCY40A and FCY20A, which have different Al$_{2}$O$_{3}$ contents and they are 100, 40 and 20, respectively, as shown by open symbols in Fig. \ref{fig1}.
FCA10 shows the highest transmission among the examined ceramics and also shows little frequency dependence.
However, the transmittance is decreased as the Al$_{2}$O$_{3}$ content is decreased.
Moreover, the decrease of the transmittance with the frequency is remarkable as shown in Fig. \ref{fig1}, which is also apparent in the fitting parameter $b$ shown in Table \ref{tab1}.
This fact suggests that ZrO$_{2}$ has higher absorption in this frequency region than Al$_{2}$O$_{3}$.
We consider that ZrO$_{2}$ has the larger dielectric loss as compared with Al$_{2}$O$_{3}$ in this frequency region and/or the peak frequency of the dielectric loss for ZrO$_{2}$ is lower than that of Al$_{2}$O$_{3}$.
Next, from the comparison of FCY40A and Z206N, whose admixture is Y$_{2}$O$_{3}$, it was found that they have almost the same transmittance in spite of the amount of ZrO$_{2}$, where the ZrO$_{2}$ content is 60 \% for FCY20A and 95 \% for Z206N. 
Moreover, from the comparison of Z206N and Z201N, the higher transmittance in Z206N can be seen, though the ZrO$_{2}$ content in Z206N is higher than Z201N.
This fact may suggest that there exist other factors for the absorption of the electromagnetic wave in addition to the amount of ZrO$_{2}$.
They might be the admixture itself, the vacancy whose species and amount probably depend on the manufacturing process sensitively, admixtures which are not disclosed, and so on.
The transmittance of Z220, whose admixture is MgO, is extremely low as compared with that of Z206N, which has almost the same amount of ZrO$_{2}$ with Z220.
From this comparison, it is also suggested that the transmittance is very sensitive to the admixture.
It was also found that there is tradeoff tendency between the fracture toughness and the transmittance in THz region.
That is, the tougher material tends to have the lower transmittance.
The most balanced one is Z206N or FCY20A. 

Consequently, the currently used FCY20A turned out to be one of the most balanced ceramics.
Moreover, FCY20A is easier to obtain commercially.
Figure \ref{pcell} shows the cross section of the hybrid-type pressure cell for ESR measurement \cite{ref18,ref19}.
It was confirmed that the pressure can be generated up to 2.5 GPa when FCY20A is used as the inner parts material \cite{ref19}.
However, the frequency region is limited up to 400 GHz practically \cite{ref19}.
Therefore, at this moment we have to replace one or two of the inner parts with other ceramic to have better transmission to widen the frequency region at the cost of the pressure.
Figure \ref{pcell} shows a setup for the simultaneous achievement of the high pressure and the high transmission.
When two pistons are replaced by those made of FCA10, which has the highest transmission among the examined ceramics, we confirmed that the maximum pressure reaches 1.5 GPa and the frequency region can be extended up to 700 GHz successfully \cite{ref19}.
Simaltaneous achievement of the pressure up to 1.5 GPa and the frequency region up to 700 GHz has never been done in the previous systems and it is very useful to study quantum spin systems as shown in the next section.

\subsection{High-pressure ESR measurements on SrCu$_{2}$(BO$_{3}$)$_{2}$}
\begin{figure}[t]
\begin{center}
\includegraphics[width=1\linewidth]{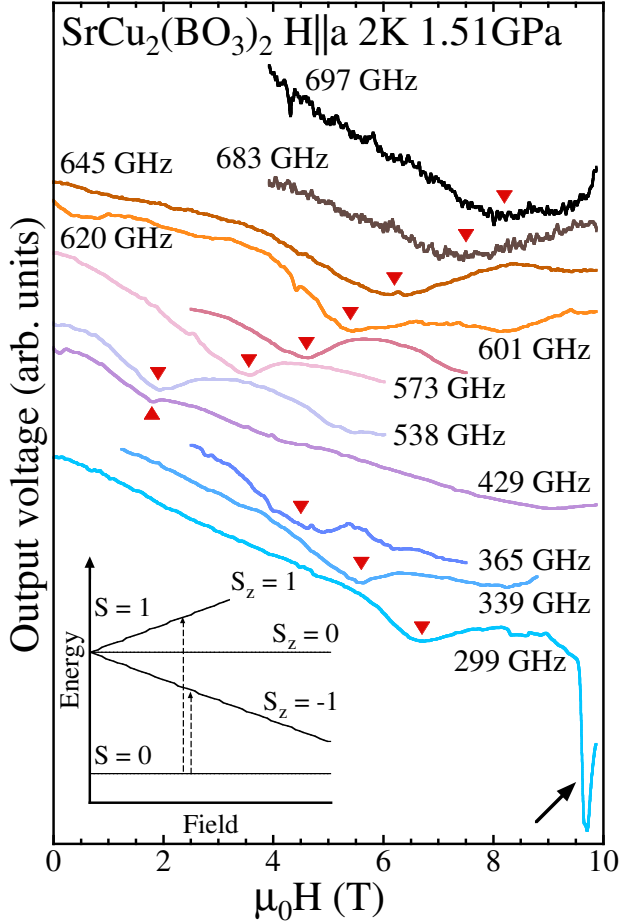}
\caption{\label{nama} Frequency dependence ESR spectra of SrCu$_{2}$(BO$_{3}$)$_{2}$ obtained at 1.51 GPa and 2 K for H$\parallel$a. The inset shows the energy-field diagram for an isolated dimer system.
Here $\mu_{0}$ is the magnetic permeability of free space.}
\end{center}
\end{figure} 

In this section we will show the result of SrCu$_{2}$(BO$_{3}$)$_{2}$ as an application of our high-pressure ESR system.
SrCu$_{2}$(BO$_{3}$)$_{2}$ has the alternate stacking of CuBO$_{3}$ layer and Sr layer along the $c$ axis.
The magnetic ion is Cu$^{2+}$ ion ($S=1/2$).
It forms the planar tetracoordinated CuO$_{4}$ with the O ions and the CuO$_{4}$ forms the Cu$_{2}$O$_{6}$ dimer by edge-sharing.
The dimers are connected orthogonally through BO$_{3}$ and they form the unique two-dimensional network in $ab$ plane. 
Figure \ref{nama} shows the frequency dependence ESR spectra obtained at the maximum pressure of 1.51 GPa in this study for the magnetic field parallel to the $a$ axis.
As is indicated by triangle symbols, the absorption lines were observed clearly in very wide frequency region.
We observed two kinds of signals, one is that shifts to the lower field side and another is that shifts to the higher field side with increasing frequency.
In the inset of Fig. \ref{nama} the energy-field diagram of an isolated dimer with the singlet ground state $S=0$ and the triplet excited states $S=1$ is shown.
The absorption lines showing the lower field side shift with the frequency and those showing the higher field side shift are assigned to the transitions from the $S=0$ to the $S_{z}=-1$ state and the $S_{z}=1$ state, respectively.
Although such direct transition between the different spin quantum number states is generally forbidden, it becomes allowed when there exists an interaction which mixes these states.
In this case, the DM interaction is considered to be responsible for this mixing \cite{nojiri2,cepas}, which is given as $\sum{\boldsymbol{D}}_{ij}\cdot\left({\boldsymbol{S}}_{i}\times{\boldsymbol{S}}_{j}\right)$, where ${\boldsymbol{D}}_{ij}$ is the DM interaction between $i$-th and $j$-th spins, and ${\boldsymbol{S}}_{i}$ is spin operator at $i$-th site.
As there is no inversion symmetry between interdimer Cu ions, the DM interaction exists between these ions.
Moreover, the ${\boldsymbol{D}}$ vector is considered to be almost along the $c$ axis from the symmetry.
The existence of the in-plane component of the interdimer DM interaction and the intradimer DM interaction can be allowed because of the small buckling of the CuO$_{4}$ plane.
However, these components are expected to be much smaller than that along the $c$ axis of the interdimer DM interaction \cite{cepas}.

At 299 GHz a sharp signal indicated by arrow was observed.
This signal has the $g$-value of about 2 and it was found that it is caused by the magnetic impurity of the Al$_{2}$O$_{3}$ inner part of the pressure cell.
When ESR signal, whose $g$-value is different from 2, is observed, the impurity signal does not overlap the ESR signal as is in this case.
However, in many cases such magnetic impurity may hide an intrinsic ESR signal.
Further research is required for a material which does not have the magnetic impurity in addition to the acceptable transmittance in THz region and the proper toughness.

\begin{figure}[t]
\begin{center}
\includegraphics[width=1\linewidth]{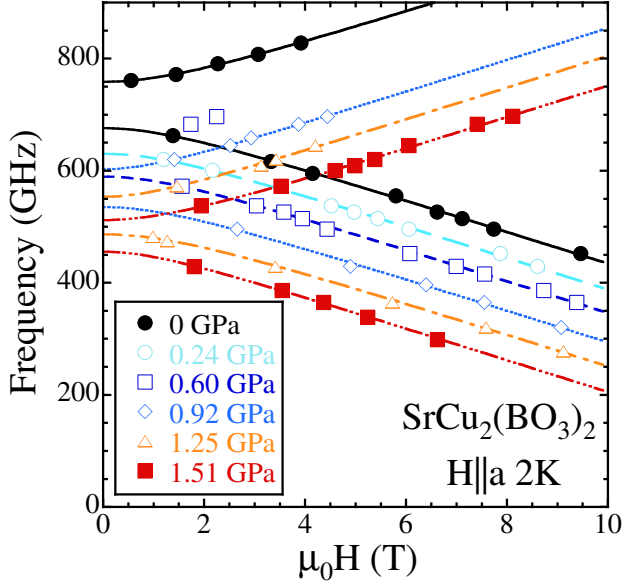}
\caption{\label{FH} Frequency-field diagram of SrCu$_{2}$(BO$_{3}$)$_{2}$ at 2 K for H$\parallel$a for various pressures.
The lines are fitting lines by the eq. (\ref{eq1}).}
\end{center}
\end{figure} 
\begin{figure}[t]
\begin{center}
\includegraphics[width=1\linewidth]{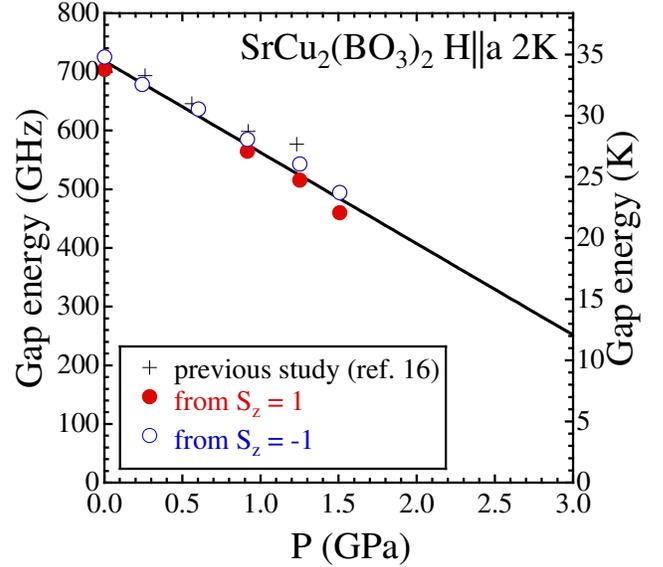}
\caption{\label{gP} Pressure dependence of the gap energy. The solid line is the result of the liner fitting to the obtained data.}
\end{center}
\end{figure} 
\begin{figure}[t]
\begin{center}
\includegraphics[width=1\linewidth]{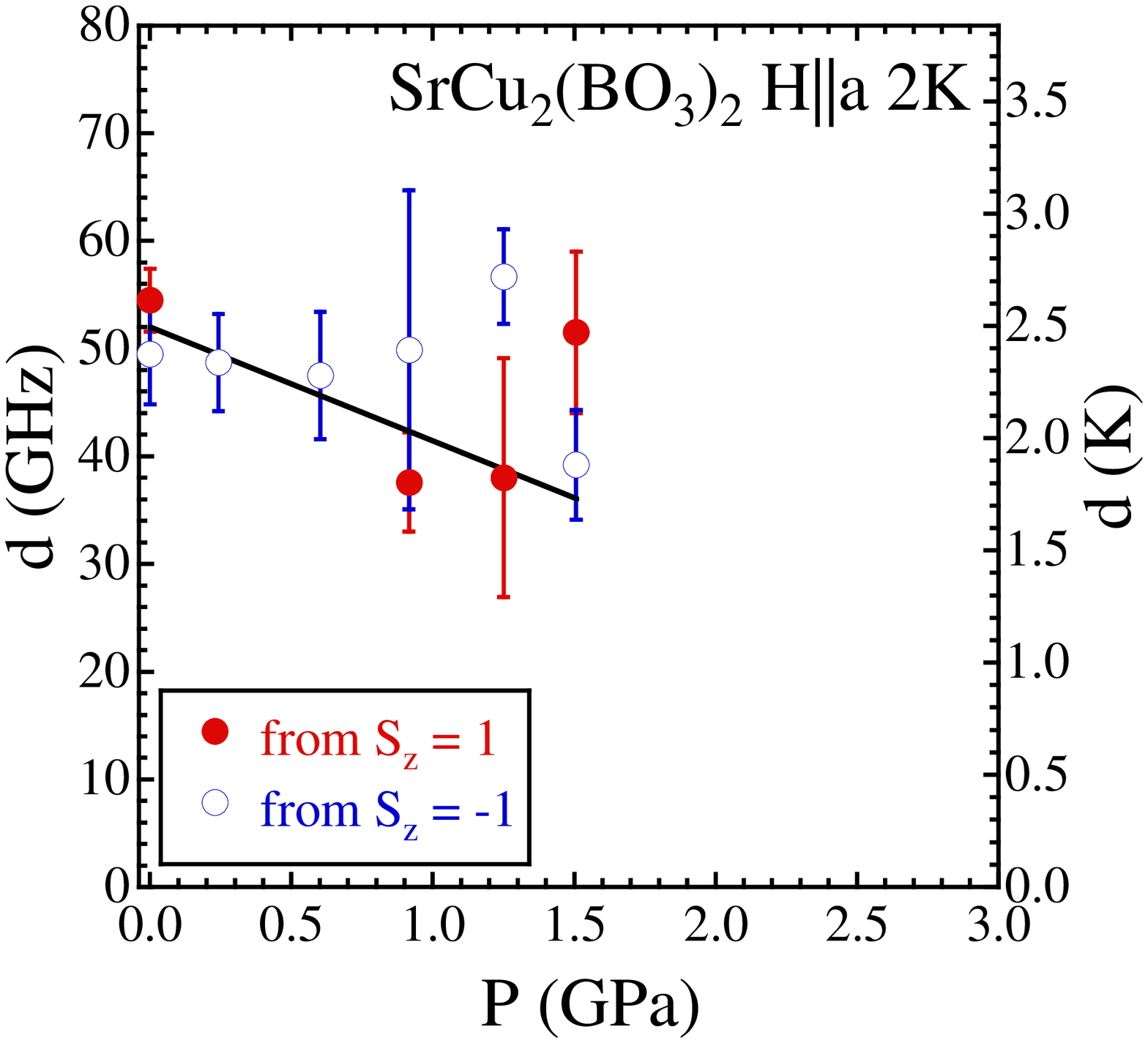}
\caption{\label{DM} Pressure dependence of the half value of the splitting between the upper and lower branches at zero field. See text for the explanation of the solid line.}
\end{center}
\end{figure} 

In Fig. \ref{FH} the resonance fields obtained at various pressures are summarized in the frequency-field diagram.
It is worth noting that such ESR mode can be obtained by only our high-pressur and multifrequency ESR system in THz region.
It is clearly shown that both the upper branch and the lower branch shift to the lower energy side as the pressure is increased. It means that the energy gap between the singlet and the triplet states is suppressed by the pressure.
It is also found that the frequency-field diagram obtained in this study differs from the simple diagram as shown in the inset of Fig. \ref{nama}.
The obtained frequency-field diagram shows the splitting at zero field and this causes the deviation from the linear field dependence of $\pm g\mu_{\rm B}H$ around zero field, where $\mu_{\rm B}$ is the Bohr magneton.
These field dependences are fitted well by the following equation \cite{cepas,room},
\begin{equation}\label{eq1}
h\nu=\Delta\pm\sqrt{d^{2}+\left(g\mu_{\rm B}H\right)^{2}}
\end{equation}
where $h$, $\nu$, $\Delta$ and $d$ are the Planck constant, frequency, gap energy and the half value of the splitting between upper and lower branch at zero field, respectively.
This equation has the same formula with that obtained theoretically by taking the DM interaction into account, in which $d$ in eq. (\ref{eq1}) can be replaced with $2D$ at the limit of $\alpha=\left(J'/J\right)\rightarrow0$, where $D$ is the magnitude of the interdimer DM interaction along the $c$ axis \cite{cepas}.
This means that the splitting at zero field is caused by the DM interaction.
However, when $\alpha$ is apart from 0 the relationship between $d$ and $D$ is no longer simple like $d=2D$ and it depends on $\alpha$.
The field range up to 10 T is not enough to evaluate the correct $g$-value and the data at the higher field side is required.
Therefore, the $g$-value is assumed to be $g=2.05$ regardless of the pressure.
This $g$-value was obtained at the paramagnetic state by Nojiri {\it et al.} and it was confirmed that it almost corresponds to that obtained from the slopes of the direct ESR transition modes in the high field \cite{nojiri}.
The curves in Fig. \ref{FH} are the fitting results.
The lines coincides with the obtained data very well.
The energy gap and the splitting at zero field obtained from this fitting are plotted as a function of pressure in Fig. \ref{gP} and \ref{DM}, respectively.
In Fig. \ref{gP} the present obtained results, the fitted line to them, and the previous result \cite{ref12} are shown.
The previous result is consistent with the data in this study.
The gap energies are obtained on average to be $\Delta_{0}=34.3$ K at ambient pressure and  $\Delta_{P}=22.9$ K at 1.51 GPa, where the subscripts 0 and $P$ indicate hereafter the values obtained at ambient pressure and 1.51 GPa, respectively.
As is clearly seen in Fig. \ref{gP}, the gap energy is reduced as the pressure is increased.
Moreover, it is found that it is reduced linearly with the pressure and there is no anomaly within the measured pressure range.
We found that the gap energy closes at 4.6 GPa from the linear extrapolation to the zero gap energy.
Figure \ref{DM} shows the pressure dependence of the splitting at zero field.
The scattering of the data is due to the lack of enough data around zero field to determine these values.

There are three unknown parameters of $J$, $\alpha$ and $D$ while we obtained only two parameters $\Delta$ and $d$ from our experiments.
Moreover, there has been no estimation of these parameters under pressure so far.
Therefore, an assumption is required to determine these parameters.
We assume that the intradimer exchange interaction $J_{0}=J_{P}=85$ K is constant under the pressure for simplicity.
This assumption seems natural because the ions are more closely packed in the Cu$_{2}$O$_{6}$ unit in which the intradimer exchange interaction works between the Cu atoms through the O atoms and this unit is much stabler than the linkage between interdimer Cu atoms.
On the other hand, the linkage between interdimer Cu atoms seems to shrink more easily by the pressure and the increase of the interdimer exchange interaction is expected under the pressure.
C\'epas {\it et al.} studied the Shastry-Sutherland lattice with the interdimer ${\boldsymbol D}$ vector of DM interaction along the $c$ axis by the exact numerical diagonalization \cite{cepas}.
They obtained the relationship between the normalized gap energy $\Delta/J$ and $\alpha$ and it shows monotonous decrease with increasing $\alpha$ up to the critical value.
That is, the decrease of the gap energy shown in Fig. \ref{gP} means the increase of the ratio $\alpha$ by the pressure under the assumption that $J$ is constant.   
Then, we can estimate $\alpha_{0}$ and $\alpha_{P}$ by comparing our results in Fig. \ref{gP} with the calculated $\alpha$ dependence of $\Delta/J$.
From our results we obtained the normalized gap energy of $\Delta_{0}/J_{0}=0.40$ at ambient pressure and $\Delta_{P}/J_{P}=0.27$ at 1.51 GPa, and the corresponding $\alpha_{0}$ and $\alpha_{P}$ are found to be $\alpha_{0}=0.64$ and $\alpha_{P}=0.68$ by reading the result of calculated $\alpha$ dependence of $\Delta/J$ in ref. 28.
As a result, the interdimer exchange interaction is increased by the pressure from $J'_{0}=54$ K at ambient pressure to $J'_{P}=58$ K at 1.51 GPa.
Moreover, it was found that the interdimer exchange interaction at ambient pressure of 54 K coincides with the theoretically obtained value completely \cite{JJ}.
We can conclude that the ratio $\alpha$ and the interdimer exchange interaction $J'$ are increased by the pressure under the assumption that the intradimer exchange interaction $J$ is constant.

C\'epas {\it et al.} also obtained the relationship between the normalized splitting at zero field  $d/D$ and $\alpha$ \cite{cepas}.
Although it is difficult to determine the proper splitting value $d$ from Fig. \ref{DM} especially in the higher pressure region because it does not have enough accuracy, at ambient pressure the splitting $d_{0}$ can be obtained relatively accurately as shown in Fig. \ref{DM}.
It is obtained as $d_{0}=2.5$ K on average.
From the calculated $\alpha$ dependence of $d/D$ at ambient pressure $\alpha_{0}=0.64$, it can be read as $d_{0}=0.91D_{0}$ \cite{cepas}.
Therefore, the DM interaction is determined to be $D_{0}=2.8$ K at ambient pressure and this value is very consistent with that obtained by C\'epas {\it et al.} $D_{0}=2.1$ K \cite{cepas}.
Since the DM interaction is considered to scale the interdimer interaction $J'$, we can evaluate the DM interaction at 1.51 GPa as $D_{P}=D_{0}\times J'_{P}/J'_{0}=2.9$ K.
From the calculated $\alpha$ dependence of $d/D$, it is also found to be $d_{P}/D_{P}=0.59$ \cite{cepas}.
Then, the value of $d_{P}$ at 1.51 GPa is expected to decrease to $d_{P}=0.59\times2.9=1.7$ K.
The change of $d$ from $d_{0}=2.5$ K at ambient pressure to $d_{P}=1.7$ K at 1.51 GPa is shown in Fig. \ref{DM} by the straight line and this line lies within the error bar.
It means that the above estimation is consistent under the assumption that the intradimer exchange interaction $J$ is constant.
In order to determine the DM interaction more directly, further precise and detailed high-pressure ESR measurement is required.
There are a few reports on the crystal structure under the pressure \cite{X-ray,X-ray2,X-ray3}.
They revealed that the lattice constant in the $ab$ plane is reduced isotropically and the structural phase transition to the monoclinic phase occurs at around 5 GPa.
The reduction of the lattice constant in the $ab$ plane agrees with the result of the increase of the interdimer exchange interaction $J'$.
However, the changes of the Cu-O-Cu angle in the Cu$_{2}$O$_{6}$ plane and the degree of the buckling of Cu$_{2}$O$_{6}$ plane under the pressure were not clarified, which are strongly correlated with the change of the intradimer exchange interaction $J$.
The investigation of these detailed structural changes are needed to verify the assumption that the intradimer exchange interaction $J$ is constant.
We found that the ratio $\alpha$ is increased by the pressure from 0.64 at ambient pressure to 0.68 at 1.51 GPa and it almost reaches the critical value of $\alpha_{c1}=0.677$ (see Fig. \ref{phase}) at this maximum pressure in this study.
However, no anomaly in ESR suggestive of the phase transition was observed at 1.51 GPa.
This fact may suggest that there are certain parameters which are not taken into account in the model Hamiltonian to obtain the phase diagram.
One of them is the exchange interaction between the layers along the $c$ axis.
From the X-ray measurement it was revealed that the decreasing rate along the $c$ axis is the highest  among three axes \cite{X-ray2}.
Although the exchange interaction between layers is small enough (8 K) at ambient pressure \cite{JJ}, the system may not be described by the simple two dimensional network as the inter layer exchange interaction is increased under the pressure.
Further theoretical study which considers the effect of the inter layer exchange interaction is required.

Finally, we comment on the X-ray measurement under pressure for this compound \cite{X-ray}.
Since we found that the system approaches the phase boundary at 1.51 GPa from the above estimation, the transition around 2 GPa suggested by the X-ray measurement is very plausible.
In contrast to this suggestion, however, it is very natural that the gap energy remains open at 2 GPa from Fig. \ref{gP}.
In the X-ray measurement, strong spin-lattice coupling is utilized to evaluate the gap energy.
That is, the lattice expansion was observed as the temperature was decreased.
Moreover, the temperature dependence of the reciprocal of the lattice constant scales the temperature dependence of the magnetic susceptibility.
Then, from the fitting to this temperature dependence of the inverse lattice constant by the simple exponential form exp$(-\Delta/kT)$ the gap energy was obtained.
This is not direct method to evaluate the gap energy and the fitting of the simple exponential form to the susceptibility is not very precise to estimate the gap energy.
On the other hand, the gap energy was directly determined in our high pressure ESR measurement and it is more reliable. 
In X-ray measurement the linear pressure dependence of the gap energy was also found but the slope (-14.7 K/GPa) \cite{X-ray} was much larger than that obtained in this study (-7.5 K/GPa).
Moreover, the above speculation that the gap energy remains open at the critical value is consistent with the suggestion that the phase transition from the dimer singlet phase to the plaquette singlet phase occurs with remaining the gap open \cite{theory2}.
The direct observation around 2 GPa by the high pressure ESR measurement is highly desired.

\section{Conclusion}
We investigated the ZrO$_{2}$-based ceramics for the inner parts of the transmission-type pressure cell used in the high-pressure and multifrequency ESR system.
The ZrO$_{2}$-Al$_{2}$O$_{3}$-based ceramics FCY20A turned out to be one of the most balanced ceramics among the examined ceramics in the toughness, transmission  property in THz region and the commercial availability.
By using the inner parts made of only Al$_{2}$O$_{3}$ ceramic which has the lowest toughness but the highest transmission together with FCY20A, we successfully reached the pressure of 1.5 GPa and extended the frequency region to 700 GHz simultaneously.

We performed the high-pressure ESR measurement of the Shastry-Sutherland compound SrCu$_{2}$(BO$_{3}$)$_{2}$.
We succeeded in observing the direct ESR transition mode between the singlet ground state and the triplet excited states up to 1.51 GPa.
The precise pressure dependence of the gap energy was obtained directly and it was proved that the gap energy is reduced by the pressure linearly.
Moreover, the deviation of the direct ESR transition mode from the linear field dependence around zero field and the splitting between the upper and lower branches at zero field are clarified by the multifrequency ESR measurement in the wide range.
From the comparison of the obtained pressure dependence of the gap energy with the theoretical result obtained by C\'epas {\it et al.} \cite{cepas}, it was found that the ratio of the exchange interactions $\alpha$ and the interdimer exchange interaction $J'$ are increased by the pressure under the assumption that the intradimer exchange interaction $J$ is constant.
Moreover, the ratio $\alpha$ reaches 0.68 at 1.51 GPa and it was suggested that the system approaches the quantum critical point.
The DM interaction was also suggested to increase by the pressure and this is consistent with the result of the pressure dependence of the splitting value at zero field.

\begin{acknowledgement}

This research was partially supported by Grants-in-Aids for Scientific Research (C) (No. 25400341) and (B) (No. 22340100) from Japan Society for the Promotion of Science.

\end{acknowledgement}



\begin{figure}[t]
\begin{center}
\includegraphics[width=1\linewidth]{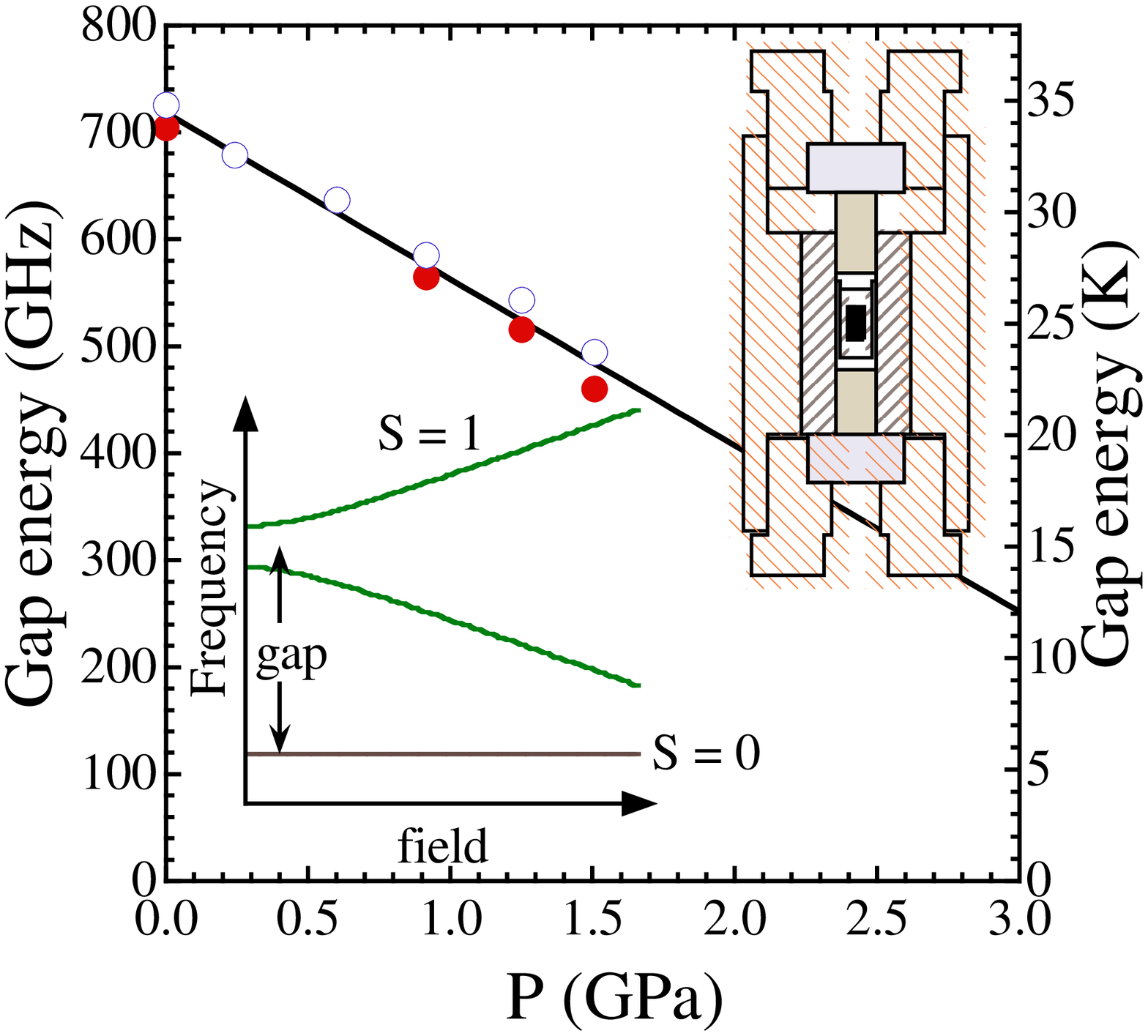}
\end{center}
\end{figure}

\end{document}